\documentclass[fleqn,twoside,twocolumn,nofootinbib,showkeys]{revtex4} % Specifies the document class %,unsortedaddress
\usepackage[sec,nocpr]{ujp} % \usepackage[cyr]{ujp} for cyrillic
\usepackage{color}
\usepackage{amssymb, graphics, setspace}
\usepackage{textcomp}
\usepackage[caption=false]{subfig}
\usepackage{verbatim}
\usepackage{amsmath}
\usepackage{commath}

\usepackage{graphicx,bm}
\usepackage{url}
\usepackage{float}
\usepackage{textcomp}
\begin{document}
\title[Pomeron-pomeron scattering]%колонтитул
{Pomeron-pomeron scattering}%
\author{Istv\'an Szanyi}%1
\affiliation{E\"otv\"os Lor\'and University}%институт 1
\address{1/A, P\'azm\'any P\'eter Walkway, 1117 Budapest, HUNGARY}%адрес 1
\email{istvan.szanyi@cern.ch}%e-mail 1
\author{Volodymyr Svintozelskyi}%2
\affiliation{\knu}%институт 2
\address{64/13, Volodymyrska Street, City of Kyiv, Ukraine, 01601 }%адрес 2
\email{1vladimirsw@gmail.com}

\udk{539} \razd{\seci}

\autorcol{I.\hspace*{0.7mm}Szanyi, V.\hspace*{0.7mm}Svintozelskyi}%

\setcounter{page}{1}%

\begin{abstract}
Central exclusive diffractive (CED) production of meson resonances potentially is a factory producing new particles, in particular a glueball. The produced resonances lie in trajectories with vacuum quantum numbers, essentially on the pomeron trajectory. A tower of resonance recurrences, the production cross section and the resonances widths are predicted. A new feature is the form of the non-linear pomeron trajectory, producing resonances (glueballs) with increasing widths. At LHC energies, in the nearly forward direction the $t$-channel both in elastic, single or double diffraction dissociation as well as in CED is dominated by pomeron exchange (the role of secondary trajectories is negligible, however a small contribution from the odderon may be present). 
\end{abstract}


\maketitle

\section{Introduction}

	Central exclusive diffractive (CED) production continues attracting attention of both theorists and experimentalists, see e.g. Ref.~\cite{FJSch} and references therein. Interest in this subject is triggered by LHC's high energies, where even the sub energies at an equal partition is sufficient to neglect the contribution from secondary Regge trajectories and consequently CED can be considered as a gluon factory to produce exotic particles such as glueballs. 

	\begin{figure*}
		\label{fig:1}
		%\vskip1mm
		\includegraphics[width=0.6\textwidth]{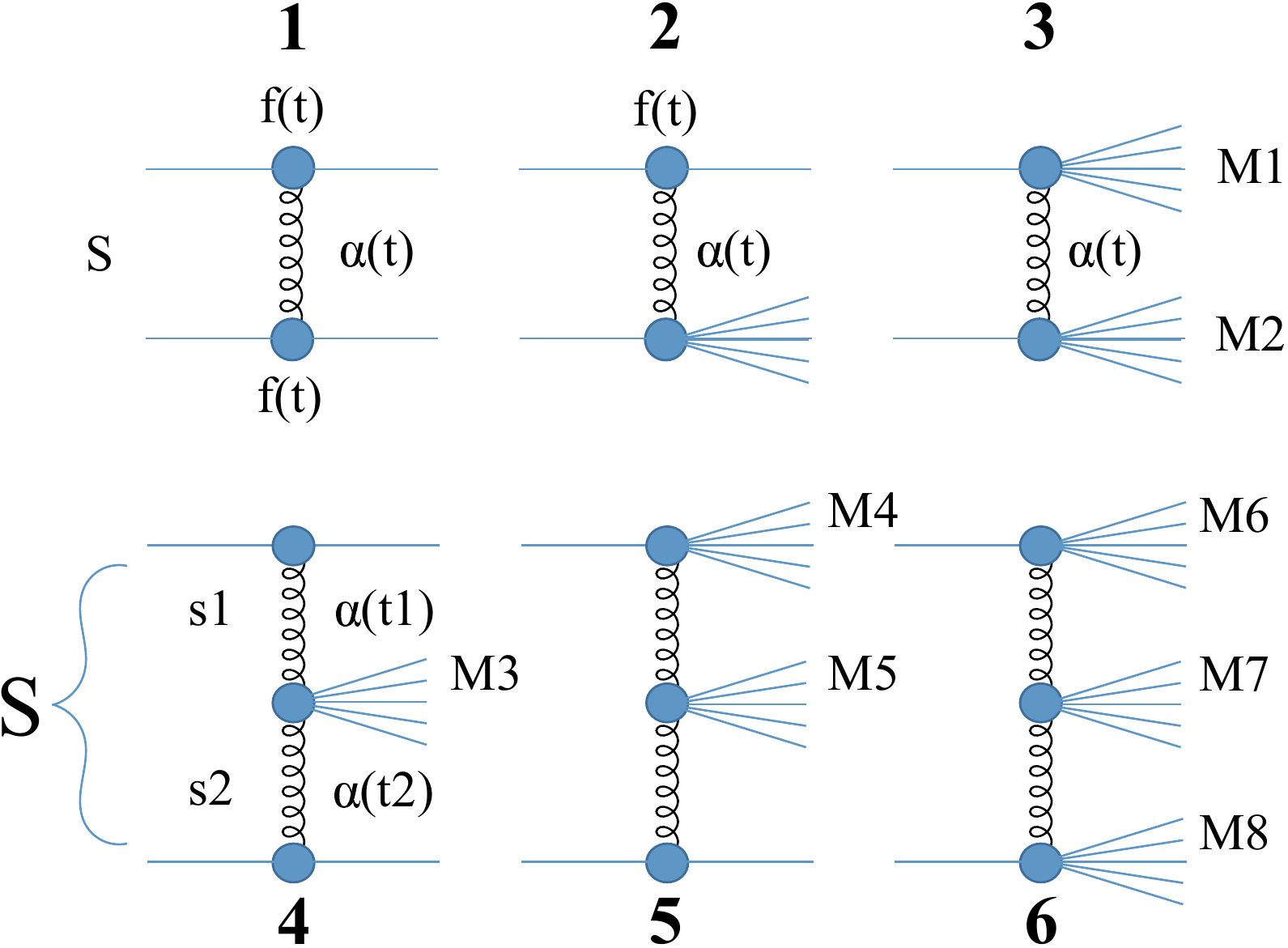}
		%\vskip-3mm
		\caption{Regge-pole factorization.}
		\label{Fig:Factor}
	\end{figure*}
	
	Below we study CED shown in Fig. \ref{fig:1} with topology 4. Its knowledge is essential in studies with diffractive excited protons, topologies 5 and 6.

	In single-diffraction dissociation or single dissociation (SD) one of the incoming protons dissociates (topology 2 in Fig.~\ref{fig:1}), in double-diffraction dissociation or double dissociation (DD) both protons dissociate (topology 3) and in central dissociation (CD) or double-Pomeron exchange (DPE) neither proton dissociates (topology 4). These processes are tabulated below as,
	\begin{eqnarray}
	{\rm SD}& pp\rightarrow Xp\nonumber\\
	{\rm or}&pp\rightarrow pY\nonumber\\
	{\rm DD}&pp\rightarrow XY\nonumber\\
	{\rm CD}\;{\rm (DPE)}&pp\rightarrow pZp,\nonumber
	\label{eqn:processes}
	\end{eqnarray}
	where $X$ and $Y$ represent diffractive dissociated protons while $Z$ denotes a central system, consisting of meson/glueball resonances.

\section{Pomeron/glueball trajectory} \label{Sec:Pomeron}

	Regge trajectories $\alpha(s)$ connect the scattering region, $s<0$, with that of particle spectroscopy, $s>0$. In this way they realize crossing symmetry and anticipate duality \textit{i.e.} dynamics of two kinematically disconnected regions is intimately related: the trajectory at $s<0$ should "know" its behavior in the cross channel and vice versa. Most of the familiar meson and baryon trajectories follow the above regularity: with their parameters fitted in the scattering region they fit masses and spins of relevant resonances, see e.g. Ref.~\cite{Collins}. The behavior of trajectories both in the scattering and particle region is close to linear, which is an approximation to reality. Resonances on real and linear trajectories imply unrealistic infinitely narrow resonances. Analyticity and unitarity also require that trajectories be non-linear complex functions \cite{Predazzi, DAMA}. Constraints on the threshold- and asymptotic behavior of Regge trajectories were derived from dual amplitudes with Mandelstam analyticity \cite{DAMA}. Accordingly, near the threshold (see also Refs.~\cite{Barut,Oehme, Gribov}) 
	\begin{equation} \label{thr}
		\Im\alpha(s)_{s \rightarrow s_0}\sim(s-s_0)^{\Re\alpha(s_0)+1/2},
	\end{equation}
	while asymptotically the trajectories are constrained by \cite{DAMA}
	\begin{equation}\label{asympt}
		\bigg\vert\frac{\alpha(s)}{\sqrt{s}\ln{s}}\bigg\vert_{s\rightarrow\infty}\leq {\rm const}.
	\end{equation}
	The above asymptotic constrain can be still lowered to a logarithm by imposing (see Ref.~\cite{Rivista} and earlier references) wide-angle power behavior for the amplitude. 
	    
	    The above constrains are restrictive but still leave much room for model building. In Refs. \cite{Francesco_m, Francesco_b} the imaginary part of the trajectories (resonances' widths) was recovered form the nearly linear real part of the trajectory by means of dispersion relations and fits to the data. 
	    
	   While the parameters of meson and baryon trajectories can be determined both from the scattering data and from the particles spectra, this is not true for the pomeron (and odderon) trajectory, known from fits to scattering data only (negative values of its argument). An obvious task is to extrapolate the pomeron trajectory from negative to positive values to predict glueball states at $J=2, 4,...$ non has been found. Given the nearly linear form of the pomeron trajectory, known from the fits to the (exponential) diffraction cone, little room is left for variations in the region of particles ($s>0$.)
	   The non-observability of any glueball state in the expected values of spin and mass may have two explanations: 1. glueballs appear as hybrid states mixed with quarks that makes their identification difficult; 2. their production cross section is low and their widths is large. To resolve these problems one needs a reliable model to predict cross sections and decay width of the expected glueballs, in which the pomeron trajectory plays a crucial role. 
	   
	   Models for the pomeron/glueball trajectories were proposed and discussed in quite a number of paper  
		\cite{1,11,11.2,11.3}. They range from simple phenomenological (also linear) models to quite sophisticated ones, involving QCD, lattice calculations, extra dimensions, etc. The basic problem of production cross section and the decay width of produced glueballs in the cited papers remains open. Close to the spirit of the present approach are papers \cite{11,11.2,11.3}, where the pomeron/glueball trajectory, including threshold singularities is manifestly non-linear and the real part terminates.

		\begin{figure}
			\vskip1mm
			\includegraphics[width=\column]{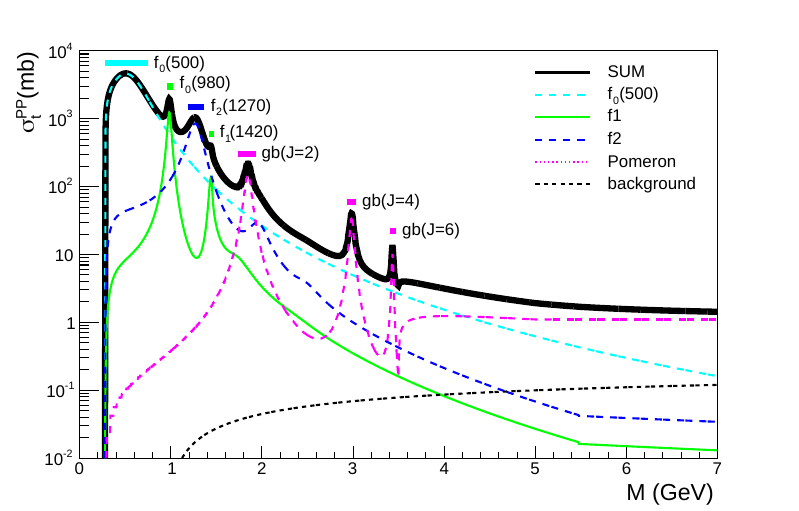}
			\vskip-3mm
			\caption{Pomeron-pomeron total cross section in CED, calculated in Ref. \cite{Schicker}.}
			\label{fig:FJSch}
		\end{figure} 

	     We continue the lines of research initiated in Refs. \cite{Schicker, FJSch} in which an analytic pomeron trajectory was used to calculate the pomeron-pomeron cross section in central exclusive production measurable in proton proton scattering e.g. at the LHC. The basic idea in that approach is the use of a non-linear complex Regge trajectory for the pomeron satisfying the requirements of the analytic $S$-matrix theory and fitting the data. Fits imply high-energy elastic proton-proton scattering data. For the scattering amplitude the simple and efficient Donnachie-Landshoff model \cite{DL} was used. The essential difference with respect to many similar studies lies in the non-linear behavior of the trajectories. They affect crucially the predicted properties of the resonances. Our previous papers \cite{Schicker, FJSch} contains more than that: the fitted trajectories are used to calculate pomeron-pomeron scattering cross sections in central exclusive diffraction at the LHC. Fig. \ref{fig:FJSch} shows the result of those calculations.   
	     
	     Papers \cite{Schicker, FJSch} contain detailed analyses and fits of both the pomeron and non-leading (also complex!) Regge trajectories, the emphases being on the pomeron/gluon one. In the present study we revise the basic object, namely the model of the pomeron trajectory, postponing other details (secondary reggeons, CED etc.) to a forthcoming study.

	\subsection{Scattering amplitude, cross sections, resonances} \label{Amplitude}

		In Ref. \cite{FJSch} the resonances contribution to pomeron-pomeron (PP) cross section was calculated from the imaginary part of the amplitude by use of the optical theorem
		\begin{equation*}
		\sigma_{t}^{PP} (M^2) \;\; = \;\; {\Im m\; A}(M^2, t=0) \;\; = 
		\end{equation*}
		\begin{equation}
		\;\; =a\sum_{i=f,P}\sum_{J}\frac{[f_{i}(0)]^{J+2}\; \Im m \;\alpha_{i}(M^{2})}
		{(J-\Re e \;\alpha_{i}(M^{2}))^{2}+(\Im m \;\alpha_{i}(M^{2}))^{2}},
		\label{eq:imampl}
		\end{equation}

		In this Section we concentrate on the pomeron. In this case Eq. (\ref{eq:imampl}) reduces to
		\begin{equation}
		\sigma_{t}^{PP} (M^2) \; = \; a\sum_{J}\frac{k^{J+2}\; \Im m \;\alpha_P(M^{2})}
		{(J-\Re e \;\alpha_P(M^{2}))^{2}+(\Im m \;\alpha_P(M^{2}))^{2}},
		\label{eq:imamplP}
		\end{equation}
		where $k=f_P(0)$, and, for simplicity here we set $k=1$.

		 We start by comparing the resulting glueball spectra in two ways: first we plot the real and imaginary parts of the trajectory (Chew-Frautchi plot) and calculate the resonances' widths by using the relation (see: e.g. Eq.~(18) in Ref.~\cite{Francesco_b})
		\begin{equation}\label{eq:Chew_F}
		\Gamma(s=M^2)=\frac{2\Im \alpha(s)}{|\alpha'(s)|},
		\end{equation}
		where $\alpha'(s)=dRe\alpha(\sqrt{s})/d\sqrt{s}$. 

		%We calculate also the ratios of the width as function of masses:
		%\begin{equation}
		%f(M)=\Gamma(M_{i+1})/\Gamma(M_i),
		%\end{equation}

	\subsection{Analytic Regge trajectories}

		In previous studies \cite{Schicker, Szanyi, FJSch} the following two types of trajectories were considered:

		\begin{equation}\label{1}
		\alpha(s)=\alpha_0+\alpha_1s+\alpha_2(\sqrt{s_0-s}-{\sqrt{s_0}}),
		\end{equation}
		and 
		\begin{equation}\label{2}
		\alpha(s)=\alpha_0+\alpha_2(\sqrt{s_0-s}-{\sqrt{s_0}})+\alpha_3(\sqrt{s_1-s}{-\sqrt{s_1}}),
		\end{equation}

		In trajectory (Eq. \ref{2}) the second, heavy threshold was introduced to mimic the nearly linear rise of the trajectory for $s<s_1$, avoiding indefinite rise as in Eq.~(\ref{1}), thus securing the asymptotic square-root upper bound Eq.~(\ref{asympt}). As realized in Refs. \cite{FJSch, Schicker}, these trajectories result in "narrowing" resonances (here, glueball) whose widths decrease as their masses increase. Below we show that this deficiency is remedied in a trajectory that satisfies the constraint of the analytic $S$ matrix theory, namely threshold behavior and asymptotic boundedness, yet produces fading resonances (glueballs), whose widths are rising with mass.    

		The trajectory is:
		\begin{equation}\label{3}
		\alpha(s)=\frac{a+bs}{1+c(\sqrt{s_0-s}-\sqrt{s_0})},
		\end{equation}
		where $s_0=4m_{\pi}^2$ and $a, b, c$ are adjustable parameters, to be fitted to scattering ($s<0$) data with the obvious constrains: $\alpha(0)\approx 1.08$ and
		$\alpha'(0)\approx {0.3}$. Trajectory Eq.~(\ref{3}) has square-root asymptotic behavior, in accord with the requirements of the analytic $S$-matrix theory.  

		With the parameters fitted in the scattering region, we continue trajectory Eq.~(\ref{3})
		to positive values of $s$. When approaching the branch cut at $s=s_0$ one has to chose the right Riemann sheet. For $s>s_0$ trajectory (\ref{3}) may be rewritten as     
		\begin{equation}\label{3m}
		\alpha(s)=\frac{a+bs}{1-c(i\sqrt{s-s_0}+\sqrt{s_0})},
		\end{equation}
		with the sign "minus" in front of $c$, according to the definition of the physical sheet. 

		For $s>>s_0,\ \  |\alpha(s)|\rightarrow\frac{b}{c}\sqrt{|s|}$. For $s>s_0$ (on the upper edge of the cut), $\Im \alpha(s)>0.$

		The intercept is $\alpha(0)=a$ and the slope at $s=0$ is 
		\begin{equation}\label{Slope}
		\alpha'(0)=b+\frac{ac}{2\sqrt{s_0}}.
		\end{equation}  

		To anticipate subsequent fits and discussions, note that the presence of the light threshold $s_0=4m_{\pi}^2$ (required by unitarity and the observed "break" in the data \cite{Tan}) results in the increasing, compared with the "standard" value of $\approx 0.25$ GeV$^{-2}$, intercept.

	\subsection{Simple Regge-pole fits to high-energy elastic scattering data}

		High-energy elastic proton-proton and proton-antiproton scattering, including ISR and LHC energies was successfully fitted with non-linear pomeron trajectories Eqs. (\ref{1}) and (\ref{2}) in number of paper, see Ref.~\cite{Tan} and references therein. 
		Since here we are interested in the parametrization of the pomeron trajectory, dominating the LHC energy region, we concentrate on the LHC data, where secondary trajectories can be completely ignored in the near forward direction. 

		While at lower energies, e.g. at the ISR, the diffraction cone shows almost perfect exponential behavior corresponding to a linear pomeron trajectory in a wide span of $0<-t<1.3$ GeV$^2$, violated only by the "break" near $t\approx -0.1$ GeV$^2$, at the LHC it is almost immediately followed by another structure, namely by the dip at $t\approx -0.6$ GeV$^2$. The dynamic of the dip (diffraction minimum) has been treated fully and successfully \cite{Szanyi}, however those details are irrelevant to the behavior of the pomeron trajectory in the resonance (positive $s$) region and expected glueballs there, that depend largely on the imaginary part of the trajectory and basically on the threshold singularity in Eq. (\ref{3}). 

\begin{figure}
			\vskip1mm
			\includegraphics[width=\column]{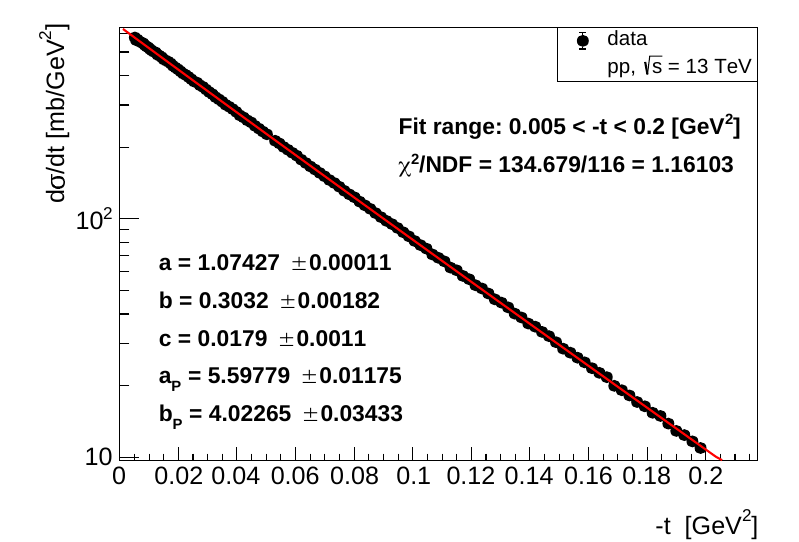}
			\vskip-3mm
			\caption{Fitted $pp$ differential cross section at 13 TeV using the amplitude Eq.~(\ref{Eq:Pf}) and trajectory Eq.~(\ref{3}).}
			\label{fig:dsdt}
		\end{figure}

\begin{figure}
			\vskip1mm
			\includegraphics[width=\column]{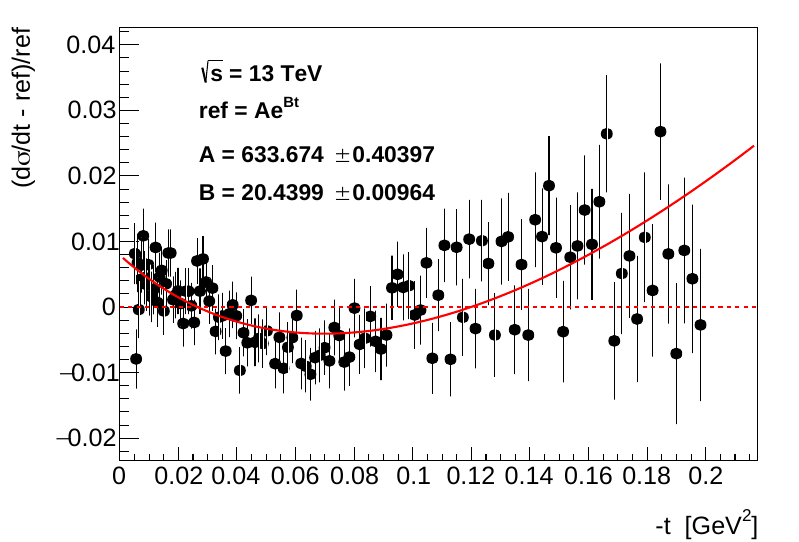}
			\vskip-3mm
			\caption{Normalized form of the fitted $pp$ differential cross section at 13 TeV using the amplitude Eq.~(\ref{Eq:Pf}) and trajectory Eq.~(\ref{3}).}
			\label{fig:dsdtnorm}
		\end{figure}

		In Fig.~\ref{fig:dsdt} we show a fit to the low-$|t|$ elastic proton-proton differential cross section data \cite{TOTEM_rho} at 13 TeV with a simple model:       
		\begin{equation}\label{Eq:Pf}
		A_P(s,t)=a_Pe^{b_Pt}e^{-i\pi\alpha_P(t)/2}(s/s_{0P})^{\alpha_P(t)},
		\end{equation}
		where $\alpha_P(t)$ is given by Eq.~(\ref{3}) (changing variable $s$ to variable $t$).
		We used the norm:
		\begin{equation}\label{Eq:Norm}
		\frac{d\sigma}{dt}=\frac{\pi}{s^2}|A_P(s,t)|^2.
		\end{equation}
Fig.~\ref{fig:dsdtnorm} shows the normalized form of the differential cross section (used by TOTEM \cite{TOTEM_rho}) illustrating the low-$|t|$ "break" phenomenon \cite{Tan} related to the non-linear square root term in the pomeron trajectory. However, it should be also noted that the "break" may result from the two-pion threshold both in the trajectory and the non-exponential residue, as discussed in Ref.~\cite{Tan}.

	\subsection{Extrapolating the pomeron trajectory to the resonance region, $s>0$}
\begin{figure}
		\vskip1mm
		\includegraphics[width=\column]{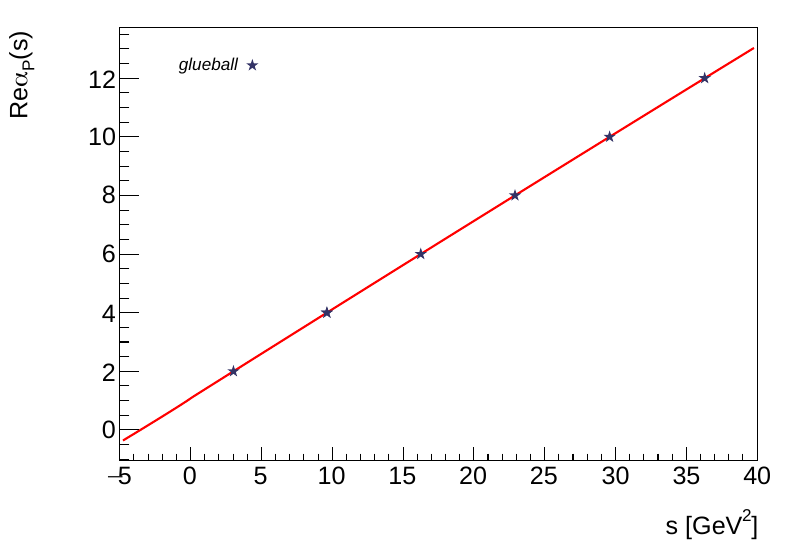}
		\vskip-3mm
		\caption{Real part of the pomeron trajectory Eq.~(\ref{3}) as function of $s$.}
		\label{fig:re_pom}
	\end{figure}

\begin{figure}
	\vskip1mm
	\includegraphics[width=\column]{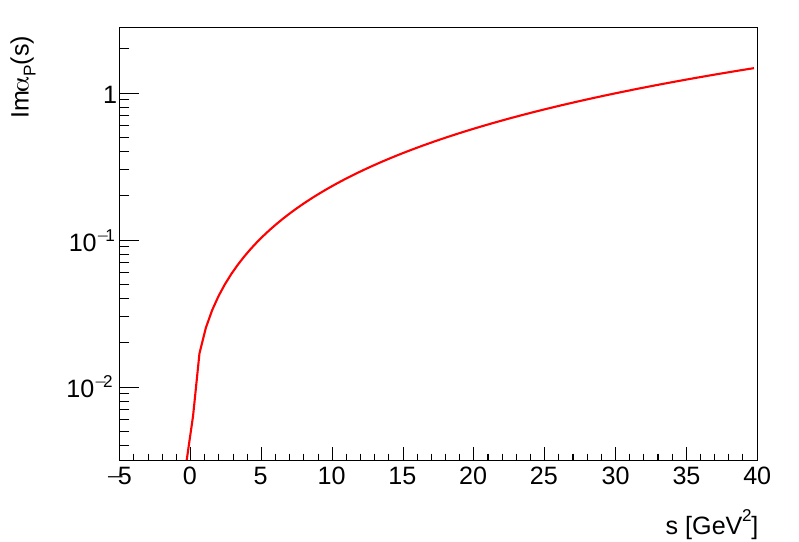}
	\vskip-3mm
	\caption{Imaginary part of the pomeron trajectory Eq.~(\ref{3}) as function of $s$. }
	\label{fig:im_pom}
\end{figure}

        Fitting to the measured $pp$ scattering data, the values of the pomeron trajectory parameters became known. Changing back variable $t$ to variable $s$ (crossing symmetry), now we can extrapolate the pomeron trajectory to the resonance region, $s>0$. Figs. \ref{fig:re_pom} and \ref{fig:im_pom} show, respectively, the real and imaginary parts of the trajectories (during the calculations, the trajectory parameter values are taken from the fit shown in Fig.~\ref{fig:dsdt}). Fig.~\ref{fig:re_pom} shows the glueball spectra lying on the pomeron trajectory. Such glueballs have even integer spins ($J\equiv\Re\alpha_{P}(s)$ = 2, 4, 6...) and mass square $M^2=s$.
		
\begin{figure}
			\vskip1mm
			\includegraphics[width=\column]{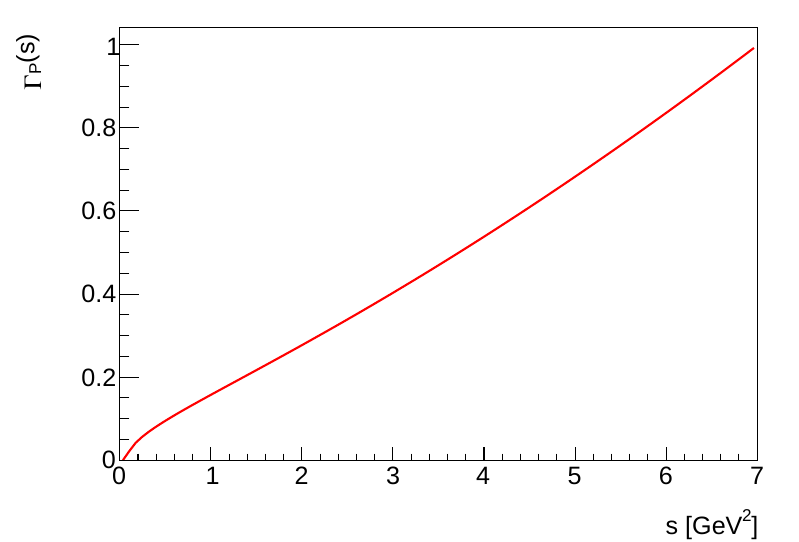}
			\vskip-3mm
			\caption{Resonance width Eq.~(\ref{eq:Chew_F}) calculated with the trajectory Eq.~(\ref{3}).}
			\label{fig:width_pom}
		\end{figure}

\begin{figure}
	\vskip1mm
	\includegraphics[width=\column]{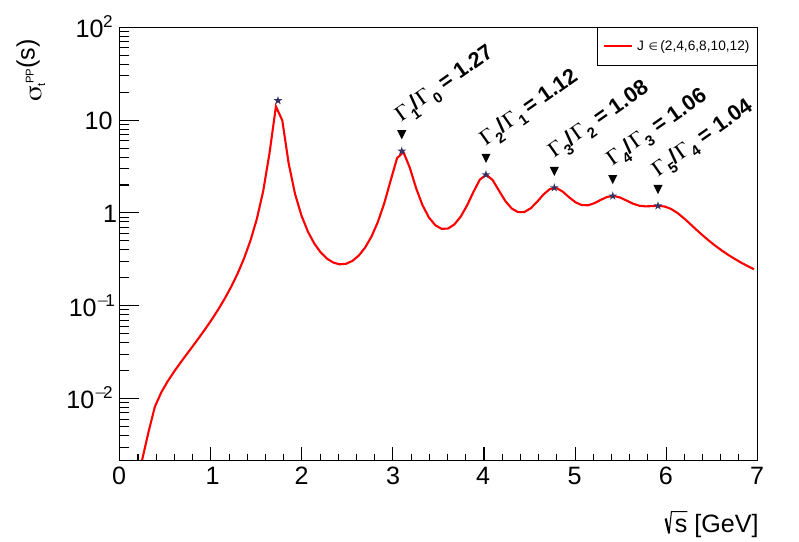}
	\vskip-3mm
	\caption{Pomeron component of the pomeron-pomeron total cross section Eq.~(\ref{eq:imamplP}) (setting $a$ = 1 GeV$^{-2}$ and $J \in (2,4,6,8,10,12)$) calculated with the trajectory Eq.~(\ref{3}). The ratios of neighboring resonances' widths are also shown.}
	\label{fig:sigma_pom}
\end{figure}

		In Fig.~\ref{fig:sigma_pom} and Fig.~\ref{fig:width_pom} we can see, respectively, the resonance width and the pomeron component of pomeron-pomeron total cross section.
	
\section{Summary}
Using a simple pomeron pole model fit to the 13 TeV $pp$ low-$|t|$ differential cross section data we have extrapolated the pomeron trajectory from negative to positive values to predict glueball states at $J=$ 2, 4, 6, 8, 10 and 12. We have predicted also the production cross section and the decay widths of the expected glueballs. Applying the pomeron trajectory Eq.~(\ref{3}) we have obtained such resonances (glueballs) whose widths increase as their masses increase. 

\vskip3mm \textit{Acknowledgment.}
We thank the organizers of the New Trends in High-Energy Physics 2019 for their support and the inspiring discussions provided by the Conference. We thank also L\'aszl\'o Jenkovszky for his guidance during the preparation of the manuscript. The work of I. Szanyi was supported by the "M\'arton \'Aron Szakkoll\'egium" program.

%\vspace*{-5mm}\rezume{%Резюме укр. мовою
%{\color{red}І. Сані}, В. Свінтозельський}{{\color{red}Розсіювання пари Померон-Померон}} {Текст резюме укр. мовою.}

\end{document}